\documentstyle[11pt,Moriond,epsfig]{article}

\newcommand{\eps}{\epsilon} 
\newcommand{\al}{\alpha}

\newcommand{\nn}{\nonumber} 
 
\newcommand{\be}{\begin{equation}} 
\newcommand{\ee}{\end{equation}} 
\newcommand{\ba}{\begin{eqnarray}} 
\newcommand{\ea}{\end{eqnarray}} 
\newcommand{\baa}{\begin{eqnarray*}} 
\newcommand{\btab}{\begin{tabular}} 
\newcommand{\etab}{\end{tabular}} 
\newcommand{\eaa}{\end{eqnarray*}} 
 
\newcommand{\bit}{\begin{itemize}} 
\newcommand{\eit}{\end{itemize}}

\newcommand\re[1]{(\ref{#1})}

\newcommand\lr[1]{{\left({#1}\right)}} 
\newcommand\lrs[1]{{\left[{#1}\right]}}

\newcommand \vev [1] {\langle{#1}\rangle}

\def \e {\mbox{e}} 
\def \CO {{\cal O}}

\def \as {\relax\ifmmode\alpha_s\else{$\alpha_s${ }}\fi}

\def \GeV {\mbox{GeV}} 
 
\begin{document} 
\begin{flushright} 
\begin{tabular}{l} 
LPT--Orsay--01--73 
\\ hep-ph/0107198 
\end{tabular} 
\end{flushright} 
\title{ Power corrections to the event shape in $e^+e^-$ annihilation. }

\author{ Sofiane Tafat  }

\address{Laboratoire de Physique Th\'eorique \def\thefootnote{\fnsymbol{footnote}}% 
\footnote{Unite Mixte de Recherche du CNRS (UMR 8627)}, 
Universit\'e de Paris XI, \\ 
91405 Orsay C\'edex, France  }

\maketitle\abstracts{We study power corrections to the event shapes in $\e^+\e^--$annihilation in the 
two jets kinematical region, $e\ll 1$. We argue that for $e \sim \Lambda_{QCD}/Q$ 
all power corrections of the form $1/\lr{Qe}^n$ have to be taken into account for 
an arbitrary $n$. This is achieved by introducing a new universal distribution, 
the shape function, which describes the energy flow in the final state. The event 
shape differential distributions are given by a convolution of the shape function 
with perturbatively resummed  cross-sections. Choosing a physically motivated 
ansatz for the shape function, we observe a good agreement of our predictions 
with available data on the differential thrust, $C$-parameter and heavy-jet mass 
distributions and their first few moments.}

%%%%%%%%%%%%%%%%%%%%%%%%%

\def\thefootnote{\arabic{footnote}} 
\setcounter{footnote} 0

\section{Introduction} 
 
The experimental data for the event shape distributions in 
$\e^+\e^--$annihilation deviate from perturbative QCD predictions by corrections 
suppressed by a power of the hard scale $1/Q^p$ with the exponent $p$ depending 
on the observable. These power corrections are associated with hadronization 
effects in the final states. Their contribution to the differential event shape 
distributions and the mean values has a different form. The leading hadronization 
corrections to the mean values can be parameterized by a single nonperturbative 
scale, whereas for the differential distribution in the end-point region, $e \sim 
\Lambda_{QCD}/Q$, one has to take into account an infinite set of the power 
corrections of the form $1/(eQ)^n$ for arbitrary $n$. Denoting by $e$ a generic 
event shape variable ($e=1-T,\rho, C$), one can write the expression for the 
differential distribution with perturbative and hadronization corrections 
included in the following general form 
$$ 
\frac{1}{\sigma_{tot}}\frac{d\sigma}{de}=\frac{d\sigma_{pert}}{de}+\frac1{Q^p}F_{hadr}(Q,e). 
$$ 
Here, $F_{hadr}(Q,e)$ receives both perturbative and nonperturbative 
contributions, which need to be disentangled in order to understand the physical 
origin of power corrections. We shall study the event shape distributions in the 
two-jet region $0 \leq e \leq e_{max}$, with the upper limit $e_{max}$ separating 
the regions with two and three jets in the final states. For the event shapes 
under consideration pQCD predictions are known to two-loop accuracy: 
\be 
\frac{d\sigma_{pert}}{de}=\frac{\alpha_s(Q)}{2\pi} A_e(e) 
\theta(e_{max}-e)+\lr{\frac{\al_s(Q)}{2\pi}}^2 B_e(e)+\CO(\al_s^3(Q)) 
\label{pQCD} 
\ee 
with $A_e(e)$ and $B_e(e)$ being coefficient functions. Due to enhancement of 
soft and collinear gluon emissions, these functions become divergent in the end 
point region, $e \to 0$, 
$$ 
A_e(e) \stackrel{e\rightarrow 0}{\sim}\: 
\frac{4C_F}{e}\left(\ln\left(\frac{e}{e_0}\right)-3/4 \right) 
$$ 
with $t_0=\rho_0=1$ and $C_0=6$. Moreover, large Sudakov logs $ \al_s^N 
\ln^{2N-n}(e)/e, \, n \geq 0 $ are present to all orders of perturbation theory. 
They spoil the convergence of perturbative expressions and need to be resummed. 
This has been done to the NLL accuracy~\cite{Catani}. It turned out that the 
resummed pQCD predictions do not fit the data in the end-point region even at the 
$Z$-mass scale (Fig.~\ref{Fig-91}a), thus calling for a better understanding of 
the underlying nonperturbative physics. 
 
\section{Event shape distributions} 
 
To understand the structure of the power corrections to the event shape 
distributions, we first have to identify the relevant scales. In the end-point 
region, the final states in the $\e^+\e^--$annihilation consist of two narrow 
quark jets surrounding by soft gluon radiation. Going through QCD analysis of 
such states, one finds the dynamics is driven by two different scales, the 
so-called soft scale, $Qe$, and collinear one, $Q\sqrt{e}$. The soft scale $Qe$ 
sets up a typical momenta of the soft gluons, while the collinear scale 
$Q\sqrt{e}$ determines the transverse size of the outgoing fast jets. 
Nonperturbative corrections to the distributions appear suppressed by powers of 
both scales. We notice, however, that in the end point region, $e\sim 
\frac{\Lambda_{QCD}}{Q}$, the following hierarchy of the scales holds, 
$Qe\ll Q\sqrt{e}\ll Q $. This suggests to neglect power corrections on the larger 
scale $Q\sqrt{e}$ and resum all corrections on the smallest scale $Qe$. The 
resulting differential distribution can be written as 
\be 
\frac{1}{\sigma_{tot}}\frac{d\sigma}{de}=\sigma_0(\alpha_s(Q),\ln(e),1/Qe)+\CO(1/Q^2e). 
\label{eq1} 
\ee 
This approximation amounts to neglecting the internal structure of two jets and 
treating them as two fast classical color charges moving in the direction of jet 
momenta and emitting soft gluons in the final states. The resulting expression 
for the differential distribution, $\sigma_0(\alpha_s(Q),\ln(e),1/Qe)$, can be 
described using Wilson loop approach ~\cite{KS-shape,KT-shape}. By the 
construction, it should resum both large perturbative Sudakov logarithms and 
power corrections on the smallest scale $Qe$ 
\begin{equation} 
\sigma_0=\frac{d\sigma_{PT}}{de}+\sum_{k=1}^{\infty}\frac{\lambda_k}{(Qe)^k}\Sigma_k(\alpha_s(Q),\ln(e)). 
\label{eq2} 
\end{equation} 
Here, $\lambda_k$ are some nonperturbative scales depending on the choice of the 
event shape variable, and $\Sigma_k(\alpha_s(Q),\ln(e))$ are perturbatively 
calculable coefficient functions. The equation (\ref{eq2}) can be thought of as a 
generalization of the OPE expansion to the event shape distributions.  For 
$e\gg{\lambda_{QCD}}/{Q}$ the leading nonperturbative corrections come only from 
the first term in the sum (\ref{eq2}). Their net effect on the distribution is 
the shift of the pQCD spectrum towards larger values of the event shape 
\cite{KS95,DMW,DW} 
$$ 
\sigma_0 
{=}\frac{1}{\sigma_{tot}}\frac{d\sigma_{pert}}{de}(e-\lambda_1/Q),\qquad 
\mbox{as $e\gg{\lambda_{QCD}}/{Q}$}. 
$$ 
The IR renormalons models give the same prediction with $\lambda_1$ being 
universal nonperturbative parameter related to the integral of the coupling 
constant over small momentum region. 
 
For $e\sim \frac{\Lambda_{QCD}}{Q}$ all terms in (\ref{eq2}) become equally 
important and need to be resummed. This is achieved  by introducing a new 
nonperturbative distribution, the so-called shape function.  Taking the thrust as 
an example we have 
\footnote{Similar formulae hold for the other event shape variables.}: 
\be 
\frac{1}{\sigma_{tot}}\frac{d\sigma}{dt} = \int_0^{Qt} d\eps f_t(\eps) 
\frac{d\sigma_t^{pert}}{dt}(t-\eps/Q). 
\label{convolution} 
\ee 
Here, $f_t(e)=\int d\eps_L d\eps_R~ f(\eps_L, \eps_R)\delta(\eps-\eps_L -\eps_R)$ 
with the shape function  $f(\eps_L,\eps_R)$ being a universal nonperturbative 
distribution describing the energy flow into the left and right hemispheres in 
$\e^+\e^--$annihilation final states. The universality of $f(\eps_L,\eps_R)$ 
implies that it does not depend on the hard scale $Q$. 
 
The shape function receives contributions of two different types 
\be 
f(\eps_L,\eps_R)=f_{\hbox{incl}}(\eps_L)f_{\hbox{incl}}(\eps_R)+\delta 
f_{\hbox{non-incl}}(\eps_L,\eps_R) 
\label{+-inv} 
\ee 
The first one is the inclusive contribution from gluons produced and decaying 
into the same hemisphere. This kind of event is partially taken into account by 
IR renormalons models. The second type of contribution corresponds to 
non-inclusive events, that is when gluon is produced in one hemisphere but decays 
into particles flowing into another hemisphere. The non-inclusive contribution 
describes the cross talk between two hemispheres and we expect that this effect 
is important for non-inclusive variables like the heavy-jet mass. 
 
\begin{figure}[h] 
\begin{center} 
\hspace*{-5mm} 
\epsfig{file=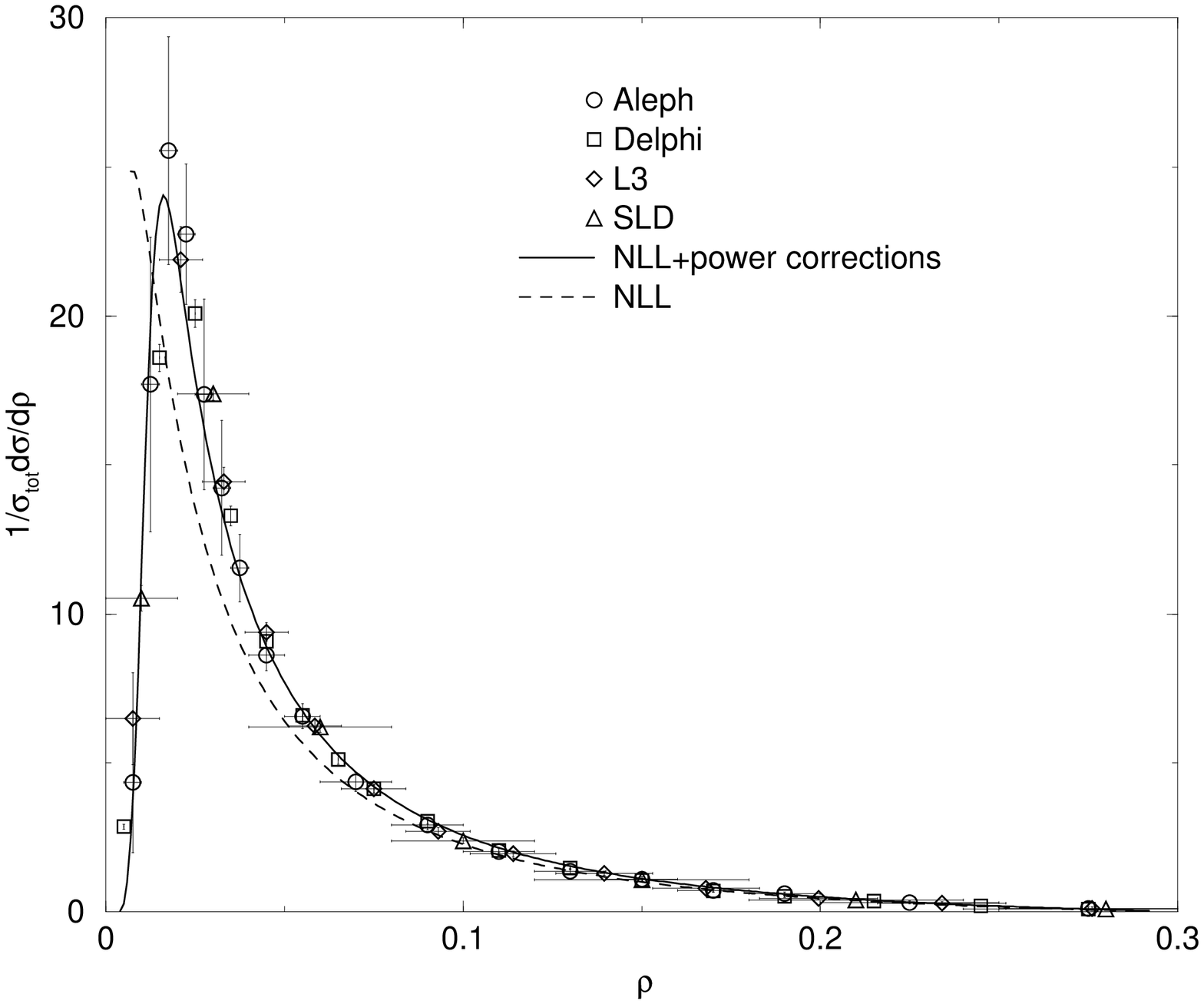,angle=-90,width=7.5cm} 
\hspace*{5mm} 
\epsfig{file=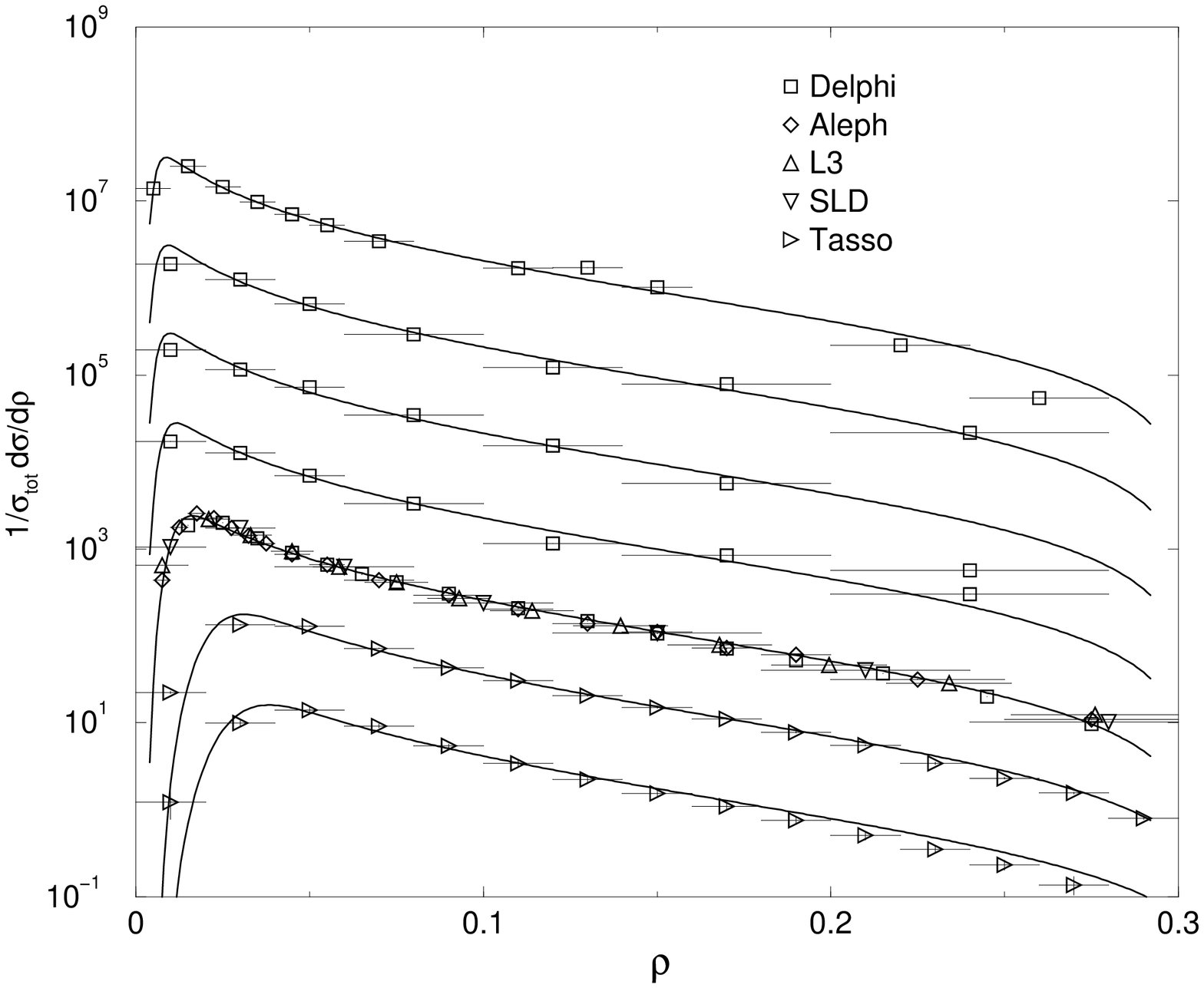,angle=-90,width=7.5cm} 
\\ 
 
{\small (a)}\hspace*{75mm}{\small (b)} 
\end{center} 
 \caption[]{(a) Heavy-jet mass distributions at $Q=M_Z$ with and without 
power corrections included. \\
(b) Comparaison of the QCD predictions for heavy jet mass at different energies from top to bottom $Q(\GeV)=35, ~44, ~91, ~133, ~161, ~172, ~183$.} 
\label{Fig-91} 
\end{figure}

The shape function is a well defined nonperturbative QCD distribution 
\cite{KS-shape,KT-shape} which in present can not be calculated from the first principles. 
Therefore, confronting our predictions with the data we shall rely on particular 
ansatz for this function. Choosing the ansatz, we require that the shape function 
should be positively definite, vanish at high energies and have a power like 
behaviour at the origin due to the phase space suppression. This leads to 
\be 
f(\epsilon_L,\epsilon_R)=\frac{{\mathcal{N}}(a,b)}{\Lambda^2} 
\left(\frac{\epsilon_L\epsilon_R}{\Lambda^2}\right)^{a-1}\exp 
\lr{-\frac{\epsilon^2_L+\epsilon_R^2+2 b \epsilon_L\epsilon_R}{\Lambda^2} }. 
\label{ansatz} 
\ee 
Here, ${\mathcal{N}}(a,b)$ is the normalization constant,  $a,~b$ and $\Lambda$ are 
nonperturbative parameters that have to be fixed by comparing our predictions for 
the event shape distributions with the data. The parameter $a$ determines how 
fast the shape function vanishes at the origin, $\Lambda$ sets up a typical scale 
of the soft radiation and $b$ controls the size of non-inclusive corrections, so 
that $\delta f_{\hbox{non-incl}}(\eps_L,\eps_R)=0$ at $b=0$. We fix the 
parameters by confronting the predictions for the heavy-jet-mass and 
$C-$parameter distributions with the data at $Q=M_Z$ (see Fig.~\ref{Fig-91}, where we present the predictions for heavy-jet mass distribution.) 
$$ 
a=2~~~~~~b=-0.4~~~~~~~\Lambda=0.55 ~\GeV. 
$$ 
To test universality of the shape function, one can use these values of the 
parameters to compare the same distributions with the data over a wide energy 
interval~\cite{exp}, $35~ \GeV~\leq Q \leq ~189 ~\GeV$. As was shown in 
\cite{KS-shape,KT-shape}, a good agreement is observed over the whole range of 
the event shapes including the end-point region $e\sim\Lambda_{\rm QCD}/Q$. 

\section{Moments} 
 
Using the obtained expressions for the differential distribution for the 
heavy-jet mass and the C-parameter, we can calculate their lowest moments. This 
calculation can be performed provided that the moments are dominated by the 
contribution of two-jet final states. One can verify that this is the case for 
the first and the second moments~\cite{G-M}. The first two moments of various 
event shape distributions have been measured by the LEP experiments~\cite{exp}. 
The obtained values deviate from the predictions of IR renormalon models, 
especially for the second moments. 
% and for variables which are not totally 
%inclusive like heavy-jet mass. 
 
Going through the calculation of the second moment of the $C$-parameter and 
heavy-jet mass distribution we find 
\begin{eqnarray} 
\vev{\rho^2}&=&\vev{\rho^2}_{PT}+\frac{\lambda_1}{Q}\vev{\rho}_{PT} 
+\frac{\lambda_2+\delta\lambda_2}{4Q^2}\nn \\ 
\vev{c^2}&=&\vev{c^2}_{PT}+\frac{3\pi}{2}\frac{\lambda_1}{Q}\lrs{2\vev{c}_{pert} 
-4.30\frac{\alpha_s(Q)}{2\pi}}+\frac{9\pi^2}{4}\frac{\lambda_2}{Q^2} 
\lrs{ 1-11.46\frac{\alpha_s(Q)}{2\pi}} 
\label{2moment} 
\end{eqnarray} 
where $\lambda_2$ is equal to the second moment of the shape function, $\lambda_2 
=\vev{\epsilon^2}=1.7 ~\GeV^2$ and $\delta \lambda_2$ measures non-inclusive 
contribution to the heavy-jet mass, 
$$ 
\delta\lambda_2 
=\vev{\left(\epsilon_L-\epsilon_R\right)^2}\left[1+4\int_0^{\rho_{max}}d\rho\, 
\rho\left(\frac{d\sigma_{PT}}{d\rho}\right)^2\right] 
$$ 
Here, $\vev{\left(\epsilon_L-\epsilon_R\right)^2}= 0.14\,\GeV^2$ and 
$\int_0^{\rho_{max}}d\rho\, \rho\left(\frac{d\sigma_{PT}}{d\rho}\right)^2$ varies 
from 2.19 for $Q=10 ~\GeV$ to 1.85 for $Q=100 ~\GeV$. We notice that non-inclusive 
correction $\delta\lambda_2$ does not affect the second moment of the 
$C-$parameter. The reason for this is that the heavy-jet mass is less inclusive 
observable with respect to a single jet than the $C-$parameter and, therefore, it 
is more sensitive to the cross-talk effects between two hemispheres. Non-inclusive 
effects provide important contribution to the second moment of the heavy-jet mass 
$\sim 14\%$. These corrections have not been taken into account in the 
IR-renormalons models, since there the two hemispheres were supposed to be 
uncorrelated. Finally, using \re{2moment} we observe a good agreement of the QCD 
predictions with the available data for the second moments of the heavy-jet and 
$C-$parameter distributions (see Fig.~\re{Fig-2nd}). 
 
\begin{figure}[] 
\begin{center}
\epsfig{file=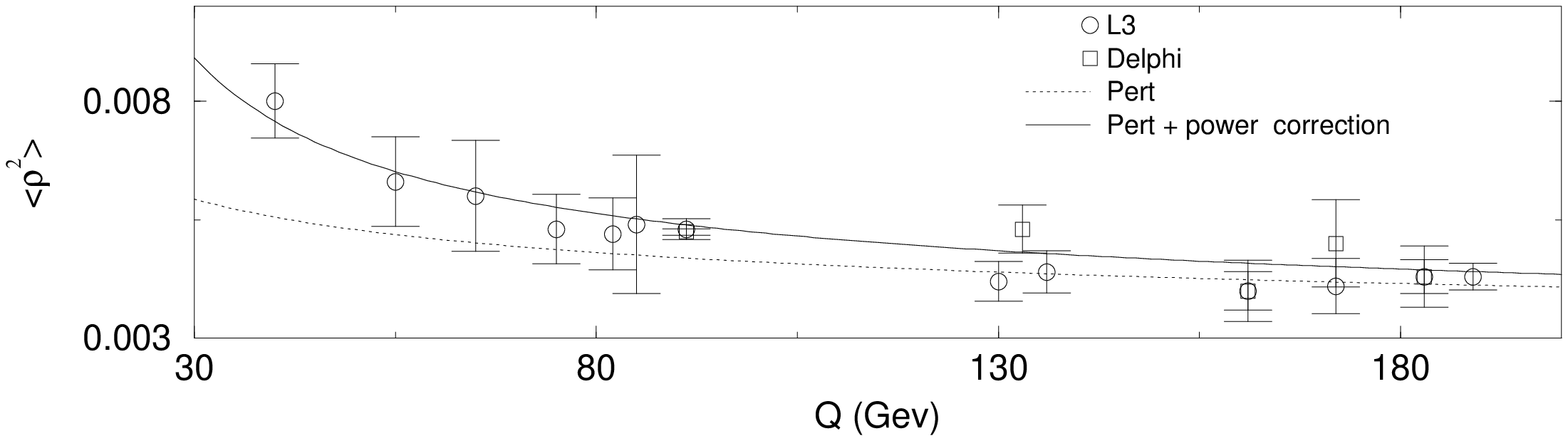,angle=-90,width=10 cm} 
\hfill 
\epsfig{file=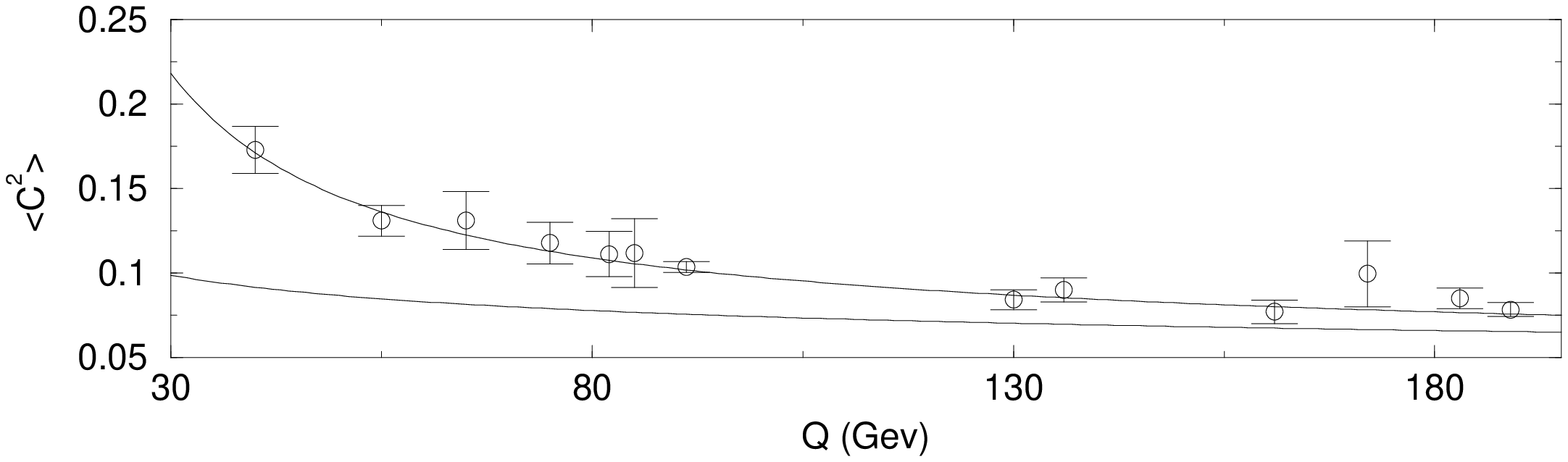,angle=-90,width=10 cm} 
\caption[]{ 
 Comparison of the QCD predictions for the second moments 
 $\vev{\rho^2}$ and $\vev{C^2}$ with the data. Dotted lines denote 
${\cal O}(\alpha_s^2)-$perturbative contribution, solid lines take into account 
power corrections given by Eq.~\ref{2moment}. } 
\label{Fig-2nd}
\end{center} 
\end{figure} 
 
\section{Conclusion} 
 
The shape function approach allows to reveal the physical meaning of the 
nonperturbative power corrections to the event shape distributions in the 
$\e^+\e^--$final states. The emerging structure of power corrections to the event 
shapes looks as follows. For $e\gg{\Lambda_{QCD}}/{Q}$ the main nonperturbative 
effect is the shift of the pQCD spectrum towards larger values of the shape 
variable, $e\to e-\lambda_e/Q$, with the scale $\lambda_e$ given by the first 
moment of the shape function. In the end point region, 
$e\sim{\Lambda_{QCD}}/{Q}$, all power corrections $\sim 1/(Qe)^n$ become equally 
important. They can be resummed in a new distribution -- the shape function which 
is a universal nonperturbative distribution describing the energy flow into two 
hemispheres in the final state. It takes into account both inclusive and 
non-inclusive corrections with the latter being especially important for the 
observable like heavy-jet mass, which are not completely inclusive with respect 
to a single jet. Choosing ansatz for the shape function, we observed a good 
agreement with the data for the differential  thrust, $C-$parameter and heavy-jet 
mass distributions as well as their mean values and the second moments.

\section*{Acknowledgments} 
I would like to thank the  Organisers of the  XXXVIth Rencontres de Moriond  for their financial support, S. Kluth for useful discussions and G. Korchemsky for  critical reading  of the manuscript.


\section*{References} 
\begin{thebibliography}{99} 
 
 
\bibitem{exp} 
O.~Biebel, 
%``Event Shapes and Power Corrections in e+e- Annihilation,'' 
hep-ex/0006020; \\ 
M.~Acciarri {\it et al.}  [L3 Collaboration], 
%``QCD studies in e+ e- annihilation from 30-GeV to 189-GeV,'' 
hep-ex/0005045; \\ 
G.~Abbiendi {\it et al.}  [JADE collaboration], 
%``QCD analyses and determinations of alpha(s) in e+ e- annihilation at  energies between 35-GeV and 189-GeV,'' 
hep-ex/0001055; \\ 
G.~Dissertori, 
%``Event shapes and power corrections in e+ e- annihilations,'' 
Nucl.\ Phys.\ Proc.\ Suppl.\  {\bf 79} (1999) 438 [hep-ex/9904033]; \\ 
P.~Abreu {\it et al.}  [DELPHI Collaboration], 
%``Energy dependence of event shapes and of alpha(s) at LEP-2,'' 
Phys.\ Lett.\  {\bf B456} (1999) 322. 
 
 
 
 
\bibitem{Catani} 
S.~Catani, L.~Trentadue, G.~Turnock and B.~R.~Webber, 
%``Resummation of large logarithms in e+ e- event shape distributions,'' 
Nucl.\ Phys.\  {\bf B407} (1993) 3. 
 
 
 
\bibitem{KS95} 
G.~P.~Korchemsky and G.~Sterman, 
%``Nonperturbative corrections in resummed cross-sections,'' 
Nucl.\ Phys.\ B {\bf 437} (1995) 415 [hep-ph/9411211]. 
 
 
\bibitem{DMW} 
Y.~L.~Dokshitzer, G.~Marchesini and B.~R.~Webber, 
%``Dispersive Approach to Power-Behaved Contributions in QCD Hard Processes,'' 
Nucl.\ Phys.\  {\bf B469} (1996) 93 [hep-ph/9512336]. 
 
\bibitem{KS-shape} 
G.~P.~Korchemsky, 
%``Shape functions and power corrections to the event shapes,'' 
in Proceedings of the 33rd Rencontres de Moriond, Les Arcs, France, 21-28 Mar 
1998, pp.~489--498 [hep-ph/9806537]; 
\\ 
G.~P.~Korchemsky and G.~Sterman, 
%``Power corrections to event shapes and factorization,'' 
Nucl.\ Phys.\  {\bf B555} (1999) 335 [hep-ph/9902341]\,. 
 
\bibitem{KT-shape} 
 
G.~P.~Korchemsky and S.~Tafat, 
%``On power corrections to the event shape distributions in QCD,'' 
JHEP {\bf 0010} (2000) 010 
[hep-ph/0007005]. 
 
\bibitem{DW} 
Y.~L.~Dokshitzer and B.~R.~Webber, 
%``Power corrections to event shape distributions,'' 
Phys.\ Lett.\  {\bf B404} (1997) 321 [hep-ph/9704298]. 
 
 
 
 
 
\bibitem{G-M} 
E.~Gardi, Talk at the 35th Rencontres de Moriond, Les Arcs, France, 18-25 Mar 
2000 [http://moriond.in2p3.fr/QCD00/transparencies/4\_wednesday/pm/gardi/]. 
 
\end{thebibliography}
\end{document}